\documentclass[12pt]{iopart}
\usepackage{iopams}
\usepackage{graphicx}
\usepackage{color}
\usepackage{amsfonts}
\input amssym.def 
\input amssym

\begin{document}

\noindent
%{\it  Trick or Truth: the Mysterious Connection Between Physics and Mathematics},\\ FQXi essay contest - Spring, 2015.

\title[]{A moonshine dialogue in mathematical physics}

\author{ Michel Planat
%$^1$
%, Alain Giorgetti$^2$, Fr\'ed\'eric Holweck$^3$ and\\ Metod Saniga$^4$ 
}

\vspace*{.1cm}
\address
{
%$^1$
 Institut FEMTO-ST, CNRS, 15 B Avenue des Montboucons, F-25033 Besan\c con, France. ({\tt michel.planat@femto-st.fr})}
%\ead{michel.planat@femto-st.fr}

\vspace*{.2cm}

\begin{abstract}\\

Phys and Math are two colleagues at the University of Sa\c{c}enbon (Crefan Kingdom), dialoguing about the remarkable efficiency of mathematics for physics. They talk about the notches on the Ishango bone, the various uses of psi in maths and physics, they arrive at dessins d'enfants, moonshine concepts, Rademacher sums and their significance in the quantum world. You should not miss their eccentric proposal of relating Bell's theorem to the Baby Monster group. Their hyperbolic polygons show a considerable singularity/cusp structure that our modern age of computers is able to capture. Henri Poincar\'e would have been happy to see it.

\end{abstract}

PACS Numbers: 00A09, 81P45, 81P13, 11G32, 20C34, 11F06, 51E12.

\vspace*{-.5cm}
%\pacs{03.65Aa, 03.65.Fd, 03.67.-a, 02.20.-a, 02.10.Ox}
%\footnotesize {~~~~~~~~~~~~~~~~~~~~~~MSC codes: 11G32, 81P13 ,81P45, 51A45,14H57, 81Q35}
%\normalsize
\section*{}
{\it Yes I'm crazy
Also they say I'm lazy
But I'll have my days
When I do as I please
You won't see the woods
While you're a tree
No you'll never see the woods
While you're a tree
} 
(Under Moonshine, The Moody Blues \cite{MoodyBlues}).

\noindent
 {\it I would suggest, as a more hopeful-looking idea for getting an improved quantum theory, that one take as basis the theory of functions of a complex variable. This branch of mathematics is of exceptional beauty, and further, the group of transformations in the complex plane, is the same as the Lorentz group governing the space-time of restricted relativity... the working out of which will be a difficult task for the future}
 (Dirac in 1939, delivered on presentation of the James Scott prize \cite{Dirac}).

\section{Day 1: the Ishango bone, psi and moonshine}
\noindent

Phys: Did you hear about the mystery of the Ishango bone found in Congo by Prof. J. de Heinzelin? You can see it pictured as Fig. 1a.

Math: Yes, this is the oldest known mathematical puzzle, it dates back at least $20,000$ years. The bone carries groups of notches totalizing $60$ in left and right columns and $48$ in the center column. A friend of mine found a good explanation of the puzzle, the ancient African of Ishango happened to use the base $12$ for counting \cite{Pletser1999}.

\begin{figure}[ht]
\centering 
\includegraphics[height=7cm]{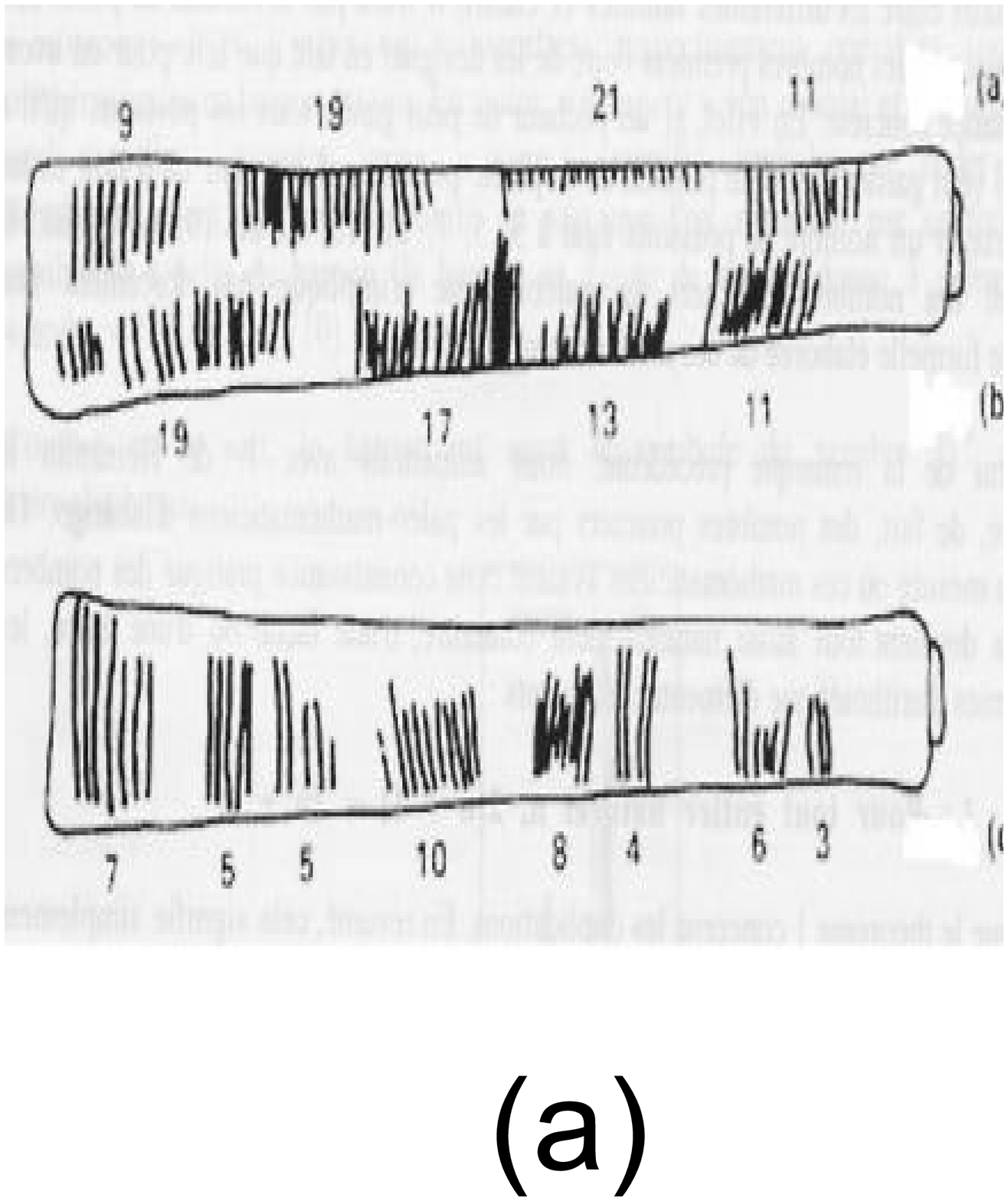}
\includegraphics[width=5cm]{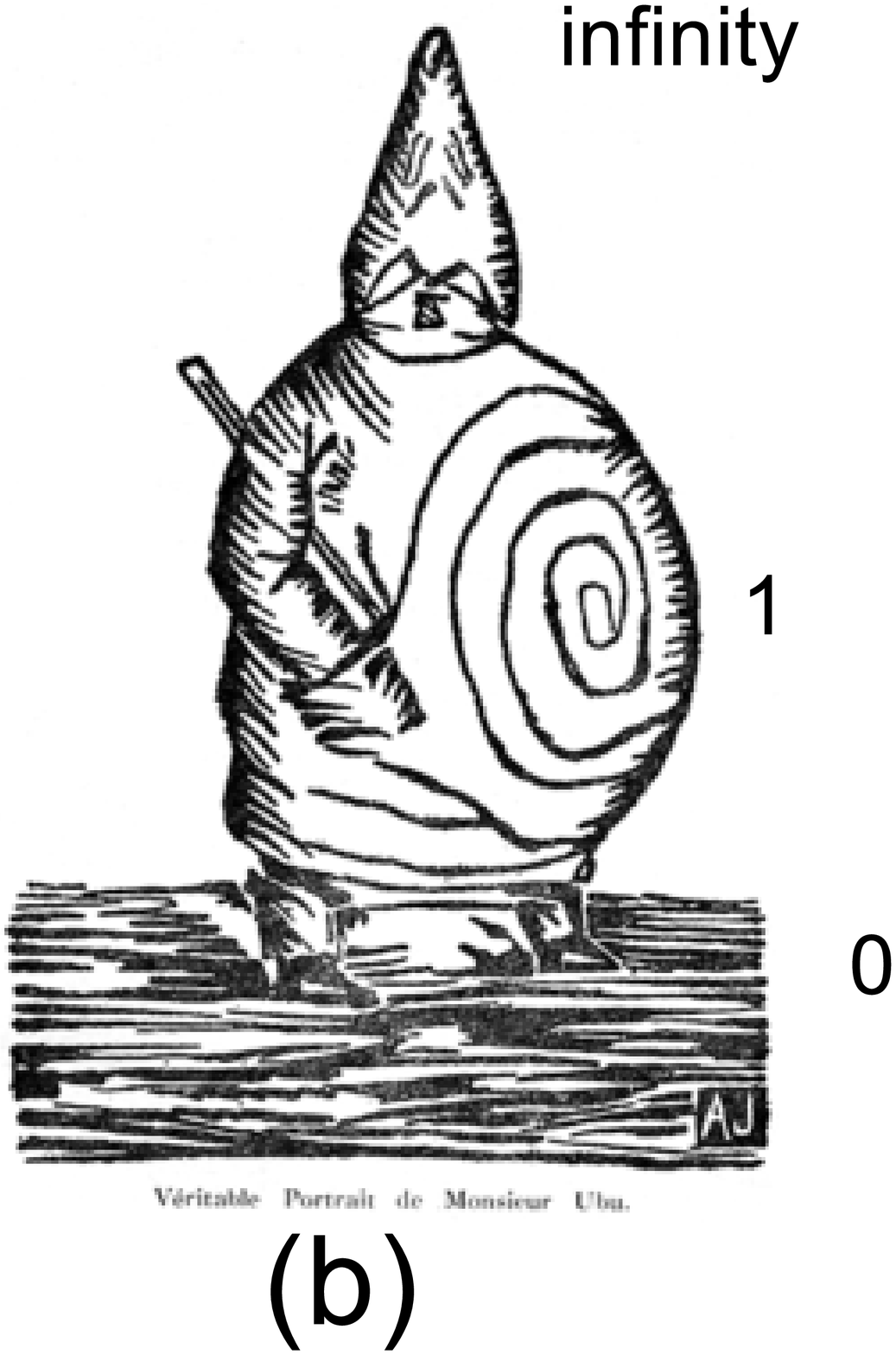}
\caption{(a) A schematic of the notches on the Ishango bone, (b) V\'eritable Portrait de Monsieur Ubu, by Alfred Jarry (1896) ( with a free labeling by the author). }
\end{figure}

Phys: Why 12?

Math: At that time, African used the thumb of a hand to count the bones in the fingers of their hands. Four fingers times three little bones on a hand yield $12$ as a counting unit. As there are 24 little bones in the two hands one gets: $60=\sigma(24)$ and $48=\psi(24)$.

Phys: What are these strange symbols $\sigma$ and $\psi$? In quantum physics, $\psi(x,t)$ denotes the wave function in space time but some scientists doubt its reality, it leads to many paradoxes such as Schr\"odinger's cat that may be simultaneously dead or alive. One finally introduced the qubit concept $\psi=a\left|0\right\rangle+b\left|1\right\rangle$ to model the superposition, but problems arise with two qubits or more with such strange phenomena as non-locality and the non-reality of objects prior to measurements, a phenomenon called contextuality.

Math: In number theory, $\psi(x)$ may designate the second Chebyshev function, a very important concept for looking accurately at the distribution of prime numbers. But I refer to the Dedekind psi function $\psi(d)=d\prod_{p|d} (1+\frac{1}{d})$, where the product is over the primes $p$ that divide $d$. At square free numbers $d$, $\psi(d)$ equals the sum of divisor function $\sigma(d)$. There are papers showing that both functions simultaneously arise for counting the number of maximal commuting sets of a $d$-level quantum system \cite{Planat2011}.

Phys: You are right. The (general) Pauli group for a qudit- a $d$-level system- is generated by two operators $X$ (shift) and $Z$ (clock) as $\mathcal{P}_d=\left\langle X,Z\right\rangle$ (generalizing the case of the two-qubit Pauli group when $X$ and $Z$ are the standard Pauli spin matrices). The operators in $\mathcal{P}_d$ are organized into $\sigma(d)$ maximal commuting sets (mcs) of size $d-1$, where $\psi(d)\le \sigma(d)$ of them are \lq admissible'. A non-admissible mcs turns out to be transversal to every admissible mcs and can be removed. The geometry of admissible mcs is the projective line $\mathbb{P}_1(\mathbb{Z}_d)$, where $\mathbb{Z}_d$ is the ring of integers $\mbox{mod}~d$, and $|\mathbb{P}_1(\mathbb{Z}_d)|=\psi(d)$. You can easily observe how much different a $4$-level system is from a $2$-qubit system. A $2$-qubit system comprises $15$ operators and $15$ mcs organized as the smallest thick generalized quadrangle $GQ(2,2)$ while a $4$-level system has the same number of operators but only $\sigma(4)=\psi(4)+1=7$ mcs \cite[Table~1]{Planat2011}.

 %I guess you mean a multidimensional Schr\"odinger's cat. This may have to do with the Cheshire cat in weak quantum measurements \cite{Denkmayr2014}. With pre- and post-selection one has the ability to separate the location of the system (the cat) from one of its properties (the grin). Cat story in Lewis Caroll tale: Alice in Wonderland is: {\it \lq\lq Well! I've often seen a cat without a grin,' thought Alice; \lq\lq but a grin without a cat! It's the most curious  thing I ever saw in my life!'} \cite{Caroll}.

Math: Both arithmetic functions $\psi(x)$ and $\psi(d)$ can be used to formulate the Riemann hypothesis (RH) about the critical zeros of the Riemann zeta $\zeta(s)$. As for the Dedekind $\psi(n)$, it was first introduced as the index of a particular congruence subgroup $\Gamma_0(q)$ of the modular group $\Gamma=\mbox{PSL}_2(\mathbb{Z})$. Congruence subgroups $\Gamma_0(p)$, for $p$ prime, are used for defining genus zero surfaces if $p-1$ divides $24$. \cite{Shimura1971}. The relation to RH is as follows. Taking the ratio $R(d)=\psi(d)/(d \log\log d)$, it has been shown that the statement $R(d)>e^{\gamma}/\zeta(2)$ (where $\gamma$ is Euler constant) is equivalent to RH \cite{SolePlanat}. As for $\psi(x)$,  I know that the logarithmic integral $\mbox{li}[\psi(x)]$ does much better than $\mbox{li}(x)$ for counting the number of primes. Here the statement $\mbox{li}[\psi(x)]-\pi(x)$ is equivalent to RH \cite{Planat2014}. 

Phys: Impressive! I am interested in advanced maths for clarifying problems in physics. I like Wolfgang Pauli quote {\it There is no God and Dirac is his prophet} \cite{Heisenberg}. Do you think that God is a mathematician or, like Alfred Jarry, that {\it God is the tangential point between zero and infinity}? \cite{Jarry}. It is a bit provocative, isn't it? So, the number $24$ occurs in the  \lq prime groups' $\Gamma_0(p)$, as it occurs in the explanation of the tracks in the Ishango bone.
It is quite remarkable that a hot topic of physics -quantum entanglement- also relates to $24$. The group of automorphisms of Euclidean dense lattices such as the root lattices $D_4$ and $E_8$, the Barnes-Wall lattice BW$_{16}$, the unimodular lattice $D_{12}^+$ and the Leech lattice $\Lambda_{24}$ may be generated by entangled quantum gates of the corresponding dimension, as shown in \cite{PhysScripta}. The Leech lattice is the densest known lattice in dimension $24$. It has the kissing number $196560$ and automorphism group $\mathbb{Z}_2.\mbox{Co}_1$, where the sporadic Conway group $\mbox{Co}_1$ has order about $4. 10^{18}$.

Math: But the story is not finished, the group you mention is a sporadic part of the largest finite group, the Monster group $\mathbb{M}$ of cardinality 
$|\mathbb{M}|=2^{46}~ 3^{20}~5^9~7^6 ~11^2~13^3~17~19~23~29~31~41~47~59~71 \sim 10^{54}$ -this would correspond to the mass in Kg of the known universe. Now if you add to $\Gamma_0(d)$ ($d$ square free) the Fricke involution matrix  $\frac{1}{\sqrt{d}}(0,-1;d,0)$
  you get another group with  a single cusp at $\infty$ called $\Gamma_0^+(d)$. For prime $d=p$, the group has genus zero if and only if $p$ is in the sequence $\{2,3,\cdots 71\}$ occurring in the factors of $[\mathbb{M}|$. This coincidence is again a puzzle when it is rewritten as 
	%\cite{Duncan2014}
	%
	\begin{eqnarray}
	&196884=1+196883,\nonumber \\
	&21493760=1+196883+21296876, \nonumber
	\end{eqnarray}
	and so on, in which the numbers at the left column occur in the $q$-expansion of the modular $j$-invariant ($q=e^{2i\pi \tau}$ and $\tau$ in the upper-half plane) and the numbers at the right column are sums of dimensions of  the smallest irreducible representations of the Monster $\mathbb{M}$ \cite{Gannon2005}. The coincidence is known as the \lq monstrous moonshine' (in the sense of being a crazy idea).  I quote 
the field medalist Richard Borcherds after his proof of the puzzle with the help of string theoretical concepts: {\it I sometimes wonder if this is the feeling you get when you take certain drugs. I don't actually know, as I have not tested this theory of mine} \cite{Borcherds}. This is an illustration of the pre-established harmony between maths and physics. 
Concerning Jarry's quote, I have put labels $0$, $1$ and $\infty$ on Ubu's portrait in Fig.~1b in order to illustrate a salient feature of Grothendieck's \lq dessins d'enfants' \cite{Groth84}, another way to approach the moonshine subject.
 
Phys: Thanks, may be this approach helps to clarify the $\psi$-quantum puzzle. Let us discuss this point tomorrow. 

\section{Day 2: the $\psi$-quantum puzzle revisited}

Phys: Look at what I see as the simplest $\psi$-quantum diagram of all, a square graph (shown in Fig. 2a) that is the paragon of two-qubit Bell's theorem about non-locality \cite[Fig. 1]{Dessins2014}. Imagine that Alice and Bob are spatially separated and do electron spin measurements (along the orthogonal directions $x$ and $z$) with their Stern-Gerlach. The four (two-qubit) operators involved are denoted $s_i$, $i=1..4$ as shown in my picture (a) where $X=(0,1;1,0)$ and $Z=(1,0;0,-1)$ are the Pauli spin matrices and a notation such as $IX$ means the tensorial product of the identity matrix $I$ and the Pauli matrix $X$. There is an edge between two vertices if they commute. As the result of a measurement can only be the eigenvalue $\pm 1$ of a $s_i$, one expects to satisfy the inequality
$$C=|\left\langle s_1s_2\right\rangle+\left\langle s_2s_3\right\rangle+\left\langle s_3s_4\right\rangle-\left\langle s_4s_2\right\rangle|\le 2.$$
But the calculations with the operators $s_i$ lead to the norm $||C||=2\sqrt{2}$ instead of $2$, and the experiments confirm this fact. There are many choices for a square graph with multiple qudits $s_i$ and the result is always a maximal violation $2\sqrt{2}$.

Math: Do you have an explanation of the algebraic equation $C^2=8$?

\begin{figure}[ht]
\centering 
\includegraphics[height=5cm]{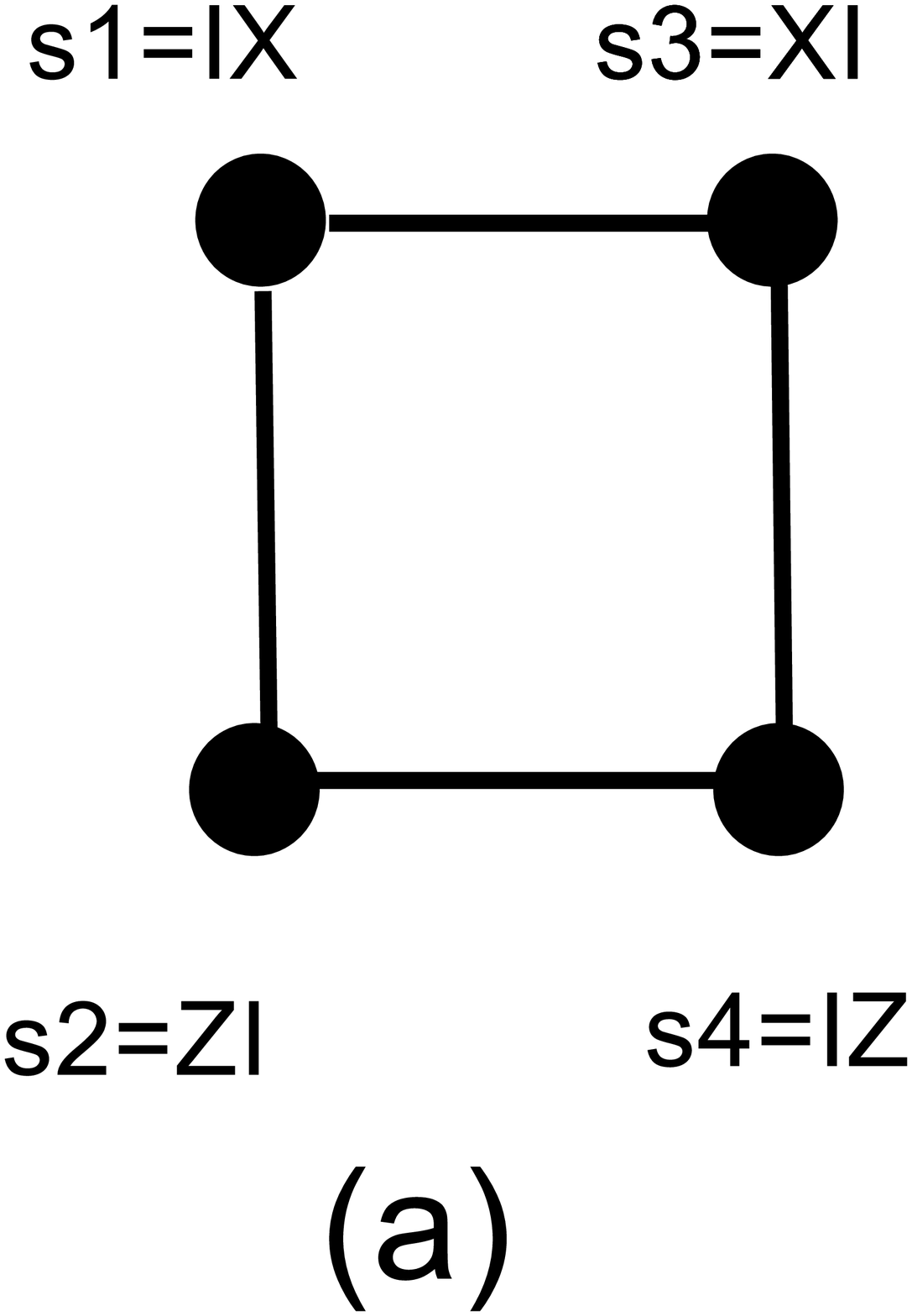}
\includegraphics[height=5cm]{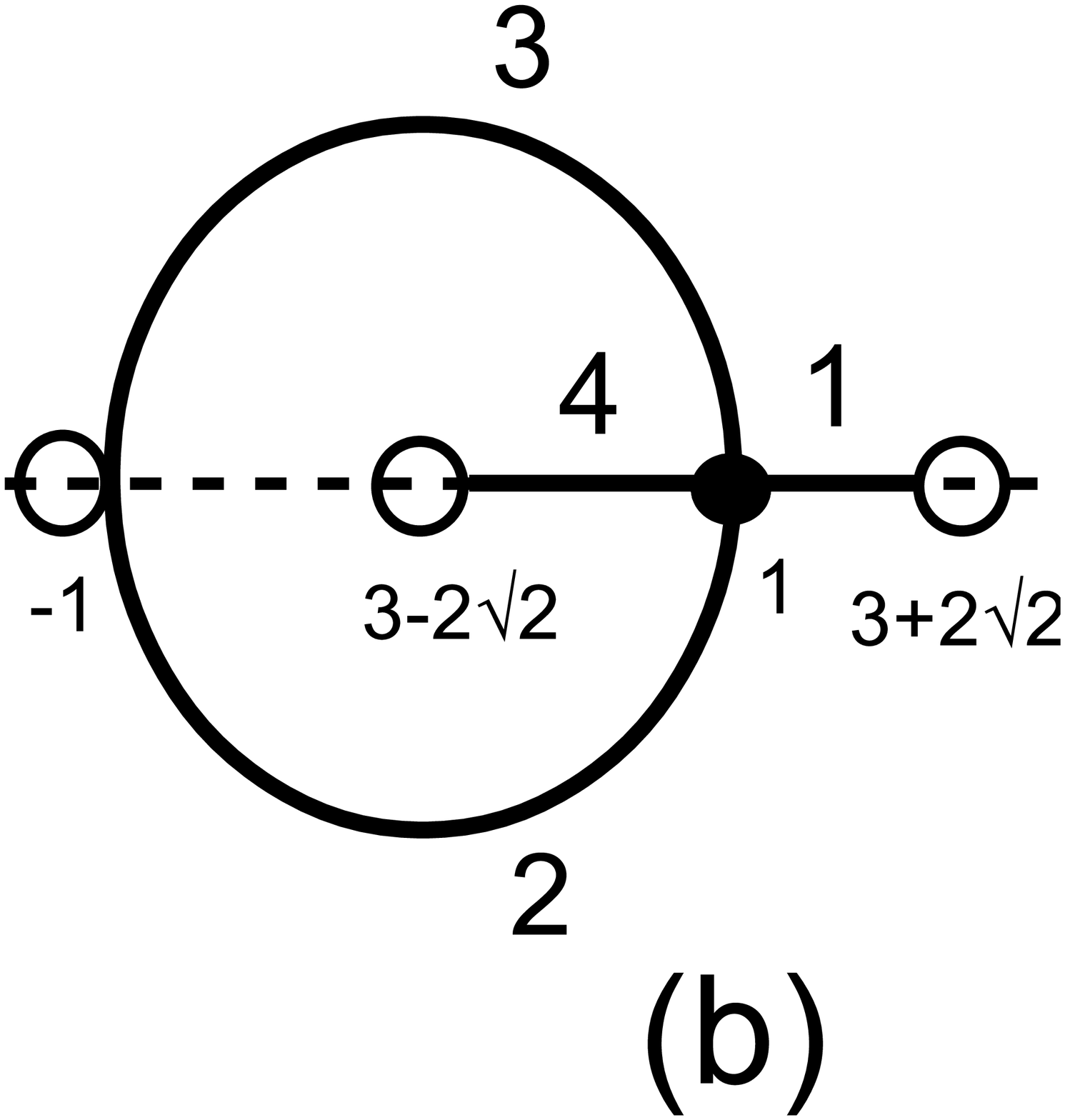}
\caption{(a) The square graph of Bell's theorem, (b) the dessin stabilizing the square graph where the labeling  \lq i' of the edges corresponds to the (operator) vertices \lq$s_i$' of the square. }
\end{figure}

Phys: One cannot escape the quantum formalism on this matter. Observe that there is none commuting entangled pair $(s_i,s_j)$ in my example, the proof of Bell's theorem does not refer to entanglement. But I found another argument based on the permutation group 
$P=\left\langle \alpha,\beta\right\rangle$ with the two generators $\alpha=(1,2,4,3)$ and $\beta=(2,3)$, this is pictured in my Fig. 2b where $i$ means $s_i$. Each edge of my drawing corresponds to the same stabilizer subgroup of $P$.

Math: Your graph is a Grothendieck's \lq dessin d'enfant' and the generators mean how you go around the black and white vertices, isn't it? And I see that the vertices live in the extension field $\mathbb{Q}(\sqrt{2})$. This means that your dessin can also be seen as a complex algebraic curve over the field of algebraic numbers. Congratulations, this is a nice use of a quite sophisticated mathematical trick. Is your graph (b) the only choice to stabilize the square?

Phys: You are right, there are essentially four choices as shown in Fig. 1 of the article \cite{Dessins2014}. I kept the most asymmetric graph because I have the vague feeling that, as Einstein wrote: {\it Everything should be made as simple as possible, but not simpler} \cite{Einstein}. The three other choices are too symmetric, I feel one needs some breaking of the symmetry to allow a deeper explanation of Bell's theorem. I suspect that mathematics can help.

Math: You know, I met your dessin before. It corresponds to one (the case I) of the non-normal inclusions of triangle groups classified by D. Singerman \cite{Hoshino2010}. The monodromy permutation $P$ that you introduced is the subcover 
$$X_0(4)\rightarrow X_0^+(2)$$
of the famous Klein quartic $X(7)=X^3Y+Y^3Z+Z^3X$. The triangle groups corresponding to the modular curves $X_0(4)$ and $X_0^+(2)$ are the congruence subgroup $\Gamma_0(4)$ and the smallest moonshine group $\Gamma_0^+(2)$ that we discussed  yesterday. I suspect that your Bell's theorem is an elementary stone of a modular physical theory based on the Monster $\mathbb{M}$, a kind of atom. This reminds me the vortex atoms of Lord Kelvin. The modern language is knot theory and Witten developed his topological quantum field theory based on this set of ideas.

Even more intriguing, the moonshine group $\Gamma_0^+(2)$ is related to the Baby Monster group $\mathbb{B}$ of order $2^{41}~3^{13}~5^6~ 7^2~ 11~ 13~17~19~23~31~47 \sim 4. 10^{33} $ through another puzzling series \cite[p. 20]{Gannon2005}
\begin{eqnarray}
	&a_1=4372=1+4371,~~~~~~~~~~a_2=96256=1+96255,\nonumber \\&a_3=1240002=2.1+4371+96255+1139374,	\nonumber
	\end{eqnarray}
	and so on, where the numbers $a_i$ correspond to the $q$-expansion of the main modular invariant $T_{2A}$ (also called a Hauptmodul) for $\Gamma_0^+(2)$ and the numbers $1,4371,96255,1139374,\ldots$ are all dimensions of smallest irreducible representations of $\mathbb{B}$.  One has the property: $\sum_{i=1}^{24} a_i^2 ~\mbox{mod}(70)=42$ as for the $q$-expansion of $j(q)$ and that of the modular discriminant $\Delta(q)$ \cite{McKay2014}.
	
That moonshine for $\mathbb{B}$ relates to Bell's theorem is a remarkable coincidence!

Phys: The approach is the application of a simple axiomatic with only two letters $a$ and $b$ \cite{Planat2015}. Mathematically, my language is also called a free group $G=\left\langle a,b|b^2=e\right\rangle$. Elements in the group are words $u$, any combination $uu^{-1}=e$ (the neutral element) is annihilated and I also write that $b$ is an involution that is, $b^2=e$.

\begin{figure}[ht]
\centering 
\includegraphics[height=5cm]{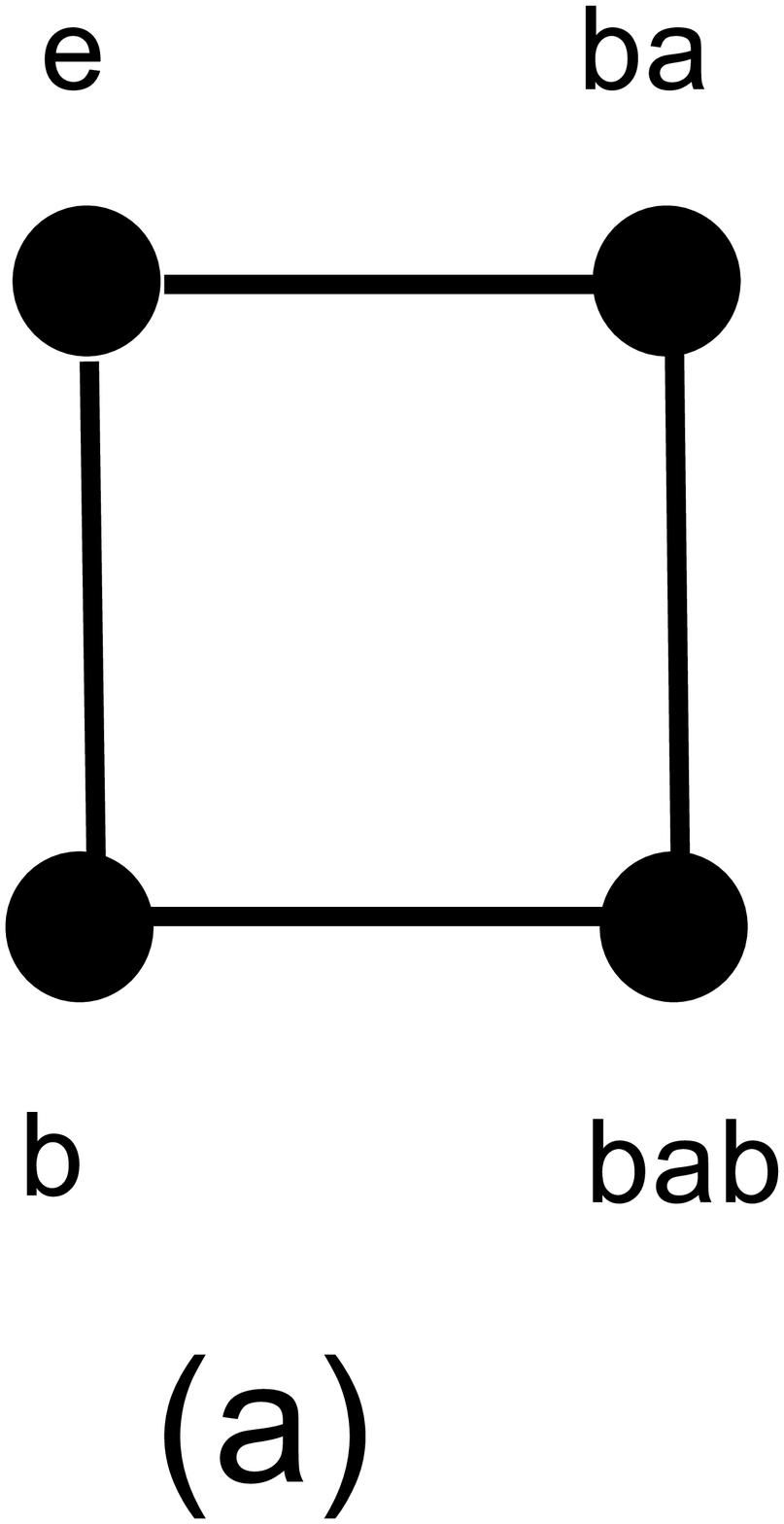}
\includegraphics[height=5cm]{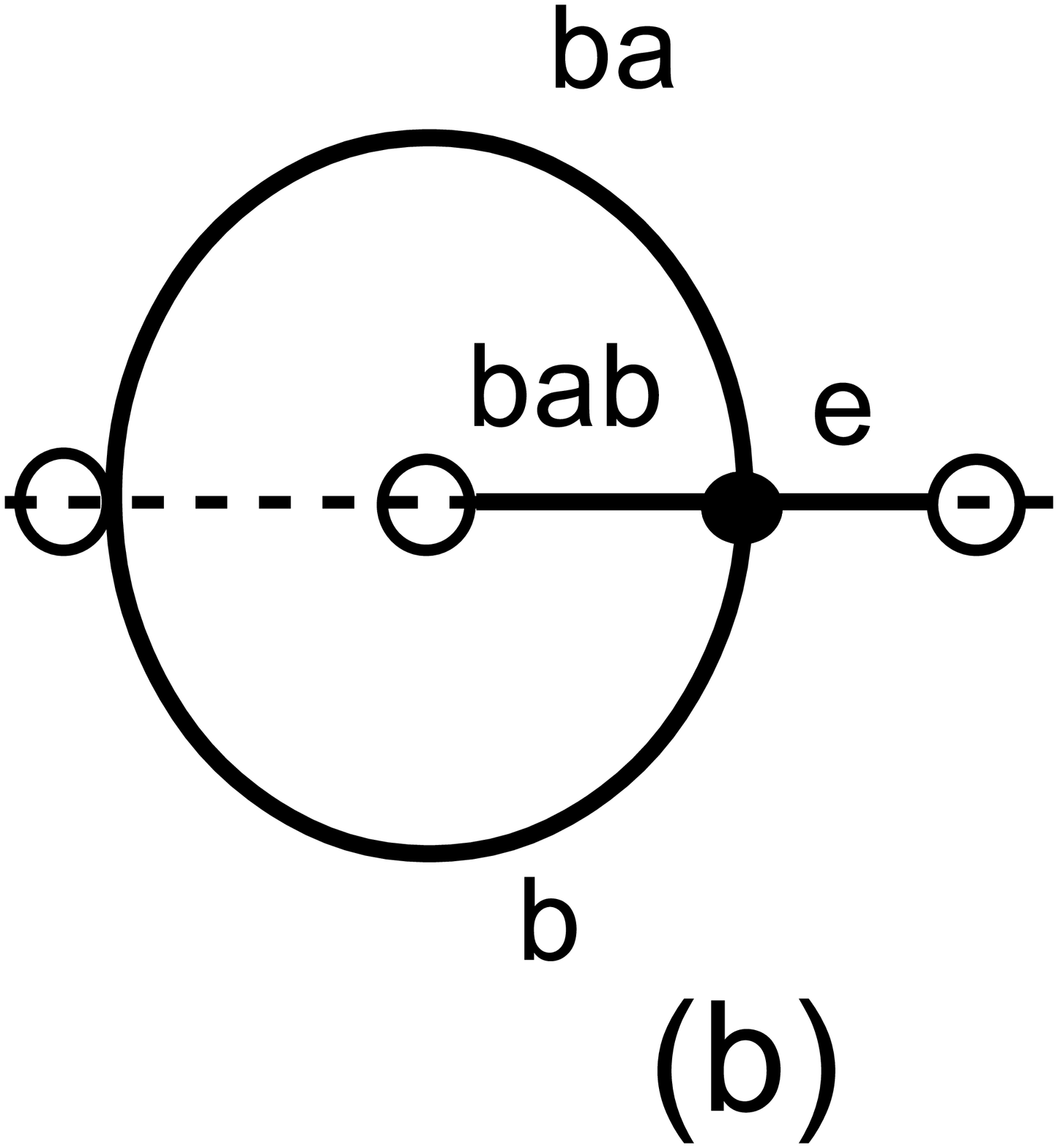}
\caption{(a) The representatives of cosets for the dessin (b) stabilizing the square graph (a). For this special case, mutually commuting operators correspond to mutually commuting cosets.}
\end{figure}

The relative size of a subgroup $H$ of $G$ is called the index which means that there are $n$ inequivalent copies (called cosets) of $H$ that fill up $G$. The action of generators on these cosets creates the permutation group $P$ by the Todd-Coxeter algorithm. You can name the cosets by a word representative, the other elements in the coset are conjugate to the representative. I did it in Fig. 3 for the case just discussed where the index is $4$. 

With a little effort, you can check that pairs of cosets on a edge are commuting in the group sense that is, the commutator $(u,v)=u^{-1}v^{-1}uv=e$. In this special case, coset commutation respects quantum commutation.

But I can show that it is not always the case for a geometry stabilized by a higher index group. Following Gottlob Frege quote: {\it Never ask for the meaning of a word in isolation, but only in the context of a sentence} \cite{Frege}.

%\caption{(a) A dessin (a) stabilizing Mermin's pentagram (b) and (c) the corresponding congruence subgroup '5C^0'. The pictures are coordinatized in terms of the $G$-set $i \in \{1..10\}$ with the corresponding cosets shown only on (a) and a set of $3$-qubit operators
%only shown on (b).}

\begin{figure}[ht]
\centering 
\includegraphics[height=7cm]{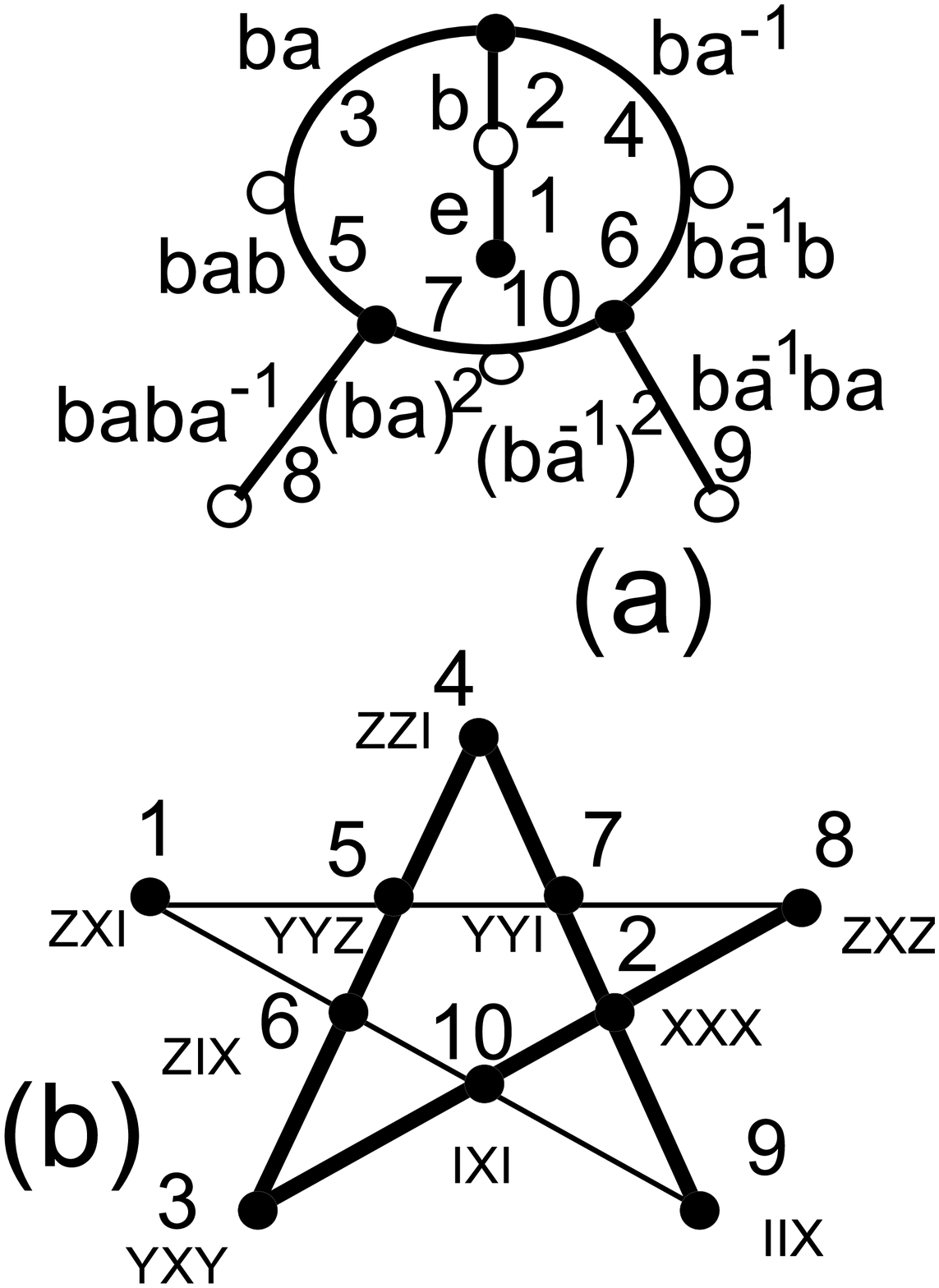}
\includegraphics[width=6cm]{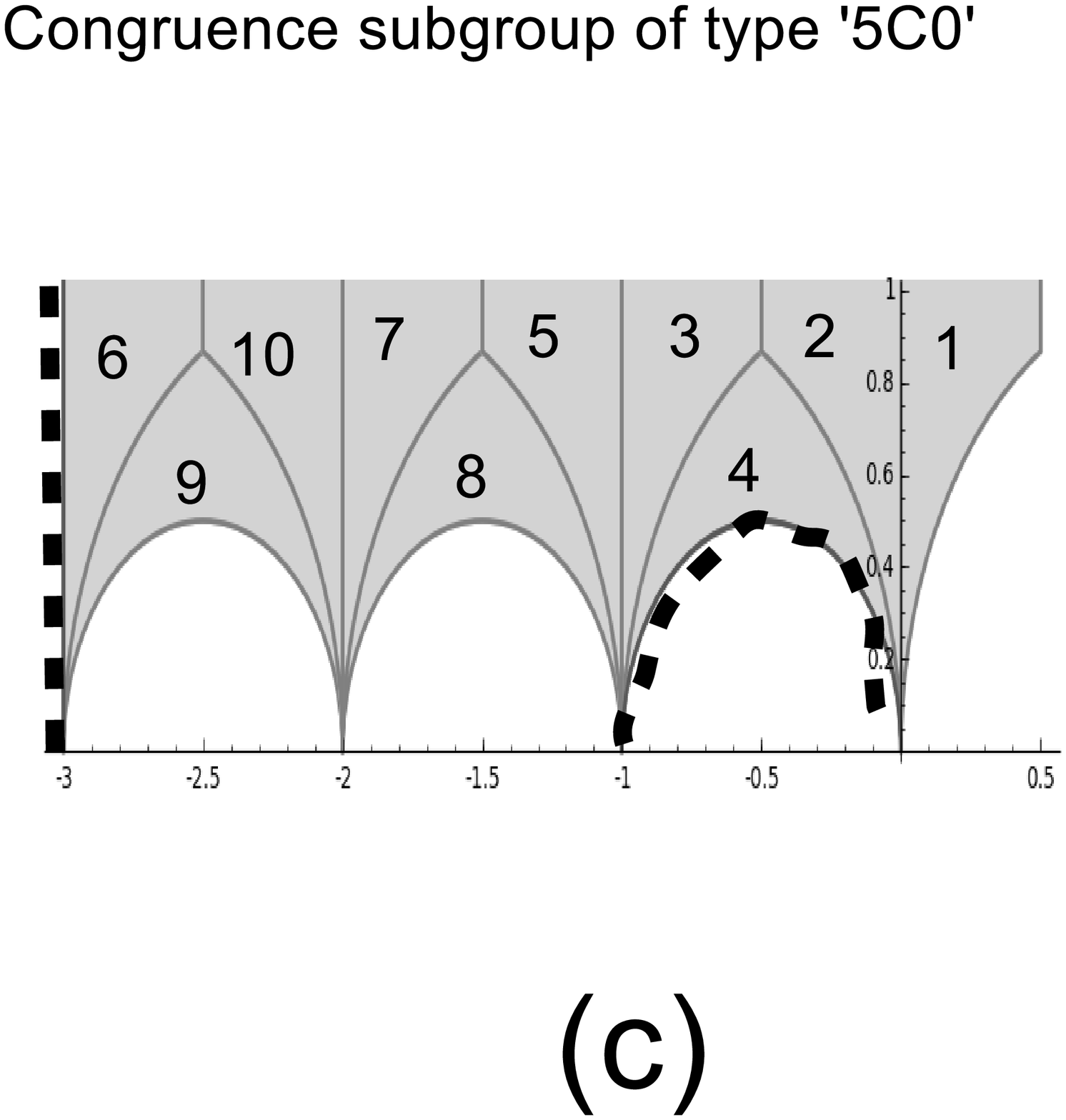}
\caption{ A dessin (a) stabilizing Mermin's pentagram (b) and (c) the fundamental domain of congruence subgroup \lq$5C^0$'. The pictures have coordinates in the $G$-set $\{1..10\}$ with the corresponding cosets shown on (a) and a set of $3$-qubit operators shown on (b).}
\end{figure}

Look at Mermin's pentagram shown on Fig. 4b \cite{Mermin1993}, I labeled the vertices  from $1$ to $10$ and with $3$-qubit coordinates. The product of operators on a thin line is $III$  and $-III$ on a thick line so that, as shown by David Mermin, this pentagram is a contextuality proof \cite{PlanatInfo,Dessins2014}.

Remember that I was able to stabilize a square with the permutation group (i.e. the dessin d'enfant) shown in Fig. 2b. With the same reasoning, I stabilize the pentagram  with a permutation group $P$ generated by the permutations $\alpha=(2,3,4)(5,7,8)(6,9,10)$ and $\beta=(1,2)(3,5)(4,6)(7,10)$ as shown in Fig. 4a. Each point on a selected line of the pentagram corresponds to the same stabilizer subgroup $S$ of $P$ (all the $S$ are isomorphic but they act on different sets of points). Then in Fig. 3b, I have put the coset representatives for labeling the edges. In this way you can check that on a thick line of the pentagram not all cosets are commuting. Of course the cosets on a (thin) line containing the identity element $e$ are commuting. I call a geometry \lq contextual' when it happens that at least a line fails to have its points/cosets commuting. In this way, geometric contextuality reflects quantum contextuality \cite{Planat2015}.

Math: You didn't comment on your Fig.~4c but I recognize the tiling of a  fundamental domain in the upper-half plane $\mathbb{H}$. I see that the tiles of $\mathbb{H}$ reflect the edge (coset) structure of your dessin and, of course, the tiles also correspond to your three-qubit operators.

Phys: Yes, the \lq modular' representation of the pentagram is in the spirit of what you explained yesterday. As the generators $\alpha$ and $\beta$ have order three and two respectively, 
they build a subgroup $\Gamma'$ of the modular group $\Gamma$ that is a congruence subgroup of level $5$. The set of cusps for 
$\Gamma'$ consists of the $\Gamma'$-orbits of $\{\mathbb{Q}\}\cup\{\infty\}$, cusps are at $-3$ and $\infty$ and they have width $3$.  My $\Gamma'$ is of type \lq $5C^0$' in Cummins-Pauli classification (http://www.uncg.edu/mat/faculty/pauli/congruence/congruence.html). I used the software Sage to draw the fundamental domain of $\Gamma'$ thanks to the Farey symbol methodology. I am not the first to play with Sage on modular aspects of dessins, you can read the essay by Lieven le Bruyn \cite{Lieven2012}. 

Math: If your approach makes sense, you should also encounter other moonshine groups, did you? 
%Do you encounter the moonshine group $\Gamma_0^+(5)$ here?

Phys: It is true that moonshine groups relate to geometric contextuality. There exists six congruence subgroups of the modular group $\Gamma$ that are simultaneously torsion free, of genus $0$ and index $12$ and three of them have a moonshine group as their normalizer \cite[Table 5]{Sebbar2002}. The corresponding dessins can be seen in \cite[Sec. 2.3.1]{He2013}. Look at the results in my Table 1: the normalizer of $\Gamma(3)$ is $\Gamma$ and three extra congruence subgroups have the normalizer $\Gamma_0 ^+(d)$, $d=2,5,6$. It is known that $j^{1/3}$ occurs as a generating function for $\Gamma(3)$ \cite[p. 7]{Gannon2005}. I also showed which geometry these groups stabilize.

\begin{table}[ht]
\begin{center}
\caption{Characteristics of torsion free, genus $0$, index $12$ congruence subgroups.}
% \vspace*{-0.4cm}
\begin{tabular}{|l|c|r|c|}
\hline \hline
Group &  Normalizer & Geometry & cusps \\
\hline\hline
$\Gamma(3)$ & $\Gamma$ &  $11$-simplex  &  $3^4$\\
$\Gamma_0(4)\cap \Gamma(2)$ & $\Gamma_0^+(2)$ &  $K(4,4,4)$ & $4^2 2^2$\\
$\Gamma_1(5)$ & $\Gamma_0^+(5)$ &  $6$-orthoplex & $5^21^2$ \\
$\Gamma_0(6)$ & $\Gamma_0^+(6)$ &  $K(6,6)$ & $6^1 3^1 2^1 1^1$ \\
$\Gamma_0(8)$ & . &  $K(4,4,4)$ & $8^2 2^1 1^2$ \\
$\Gamma_0(9)$ & . &  $K(3,3,3,3)$ & $9^1 1^3$ \\
\hline
\hline
\end{tabular}
\end{center}
\end{table}

%Not yet. There are papers about the connection of modular subgroups to dessins d'enfants, $K3$ surfaces and supersymmetric gauge theory (this is part of the so-called \lq umbral moonshine') \cite{McKay2013}. Look at such an elegant formula as $G(q)=\phi(q)^{24}$, that generates the states of a bosonic string oscillator from the Euler phi function $\phi(q)$ \cite{McKay2013} (still the number $24$ occurring).
% What about $\psi(q)^{24}$ since this is nothing but passing from the minus to the positive sign? May be this maths converts into physics after all. And the moonshine type of maths will possible clarify the quantum puzzl\section{Day 3: Rademacher sums are physical}.
Tomorrow, I tell you more about my work inspired by the moonshine topic.

\section{Dessins d'enfants, hyperbolic polygons, Rademacher sums}

Phys: It seems not to be widely known that most sporadic groups may be defined as permutation groups $P$ with two generators \cite{Mazurov1989} - acting on a $G$-set of cardinality $n$. These representations can be found explicitly in the \lq Atlas of finite group representations' \cite{Atlasv3}.  As at the previous section, a $P$ may be seen as \lq dessin d'enfant' $\mathcal{D}$ endowed with a natural topological structure with Euler characteristic $2-2 g=B+W+F-n$ with $B$ black points, $W$ white points, $F$ faces, $n$ edges and genus $g$ (see \cite{Groth84,Dessins2014,Planat2015} for details).

Another remarkable property is that, in many cases, a $P$ (and its $\mathcal{D}$), stabilizes a graph/geometry $\mathcal{G}$. First, a pair of elements of the $G$-set attached to $P$ defines a stabilizer subgroup $S$ of $P$. Second, the collection of isomorphic $S$ acting on different subsets of the $G$-set defines the edges/lines of $\mathcal{G}$. A list of the stabilized $\mathcal{G}'s$ of small index can be found in \cite[Tables 1 and 2]{Dessins2014}. In the following I restrict to \lq modular dessins' that are defined from a subgroup $\Gamma'$ of the modular group $\Gamma$ (as for the pentagram of the previous section). For these $\mathcal{D}$ the valency of black and white points is three and two, respectively (except for the elliptic points where the valency is one).  

Lieven le Bruyn writes in \cite{Lieven2012}:
{\it It would be nice to have (a) other Farey-symbols associated to the second Janko group, hopefully showing a pattern that one can extend into an infinite family as in the iguanodon series and (b) to determine Farey-symbols of more sporadic groups}. 

\begin{figure}[ht]
\centering 
\includegraphics[width=5cm]{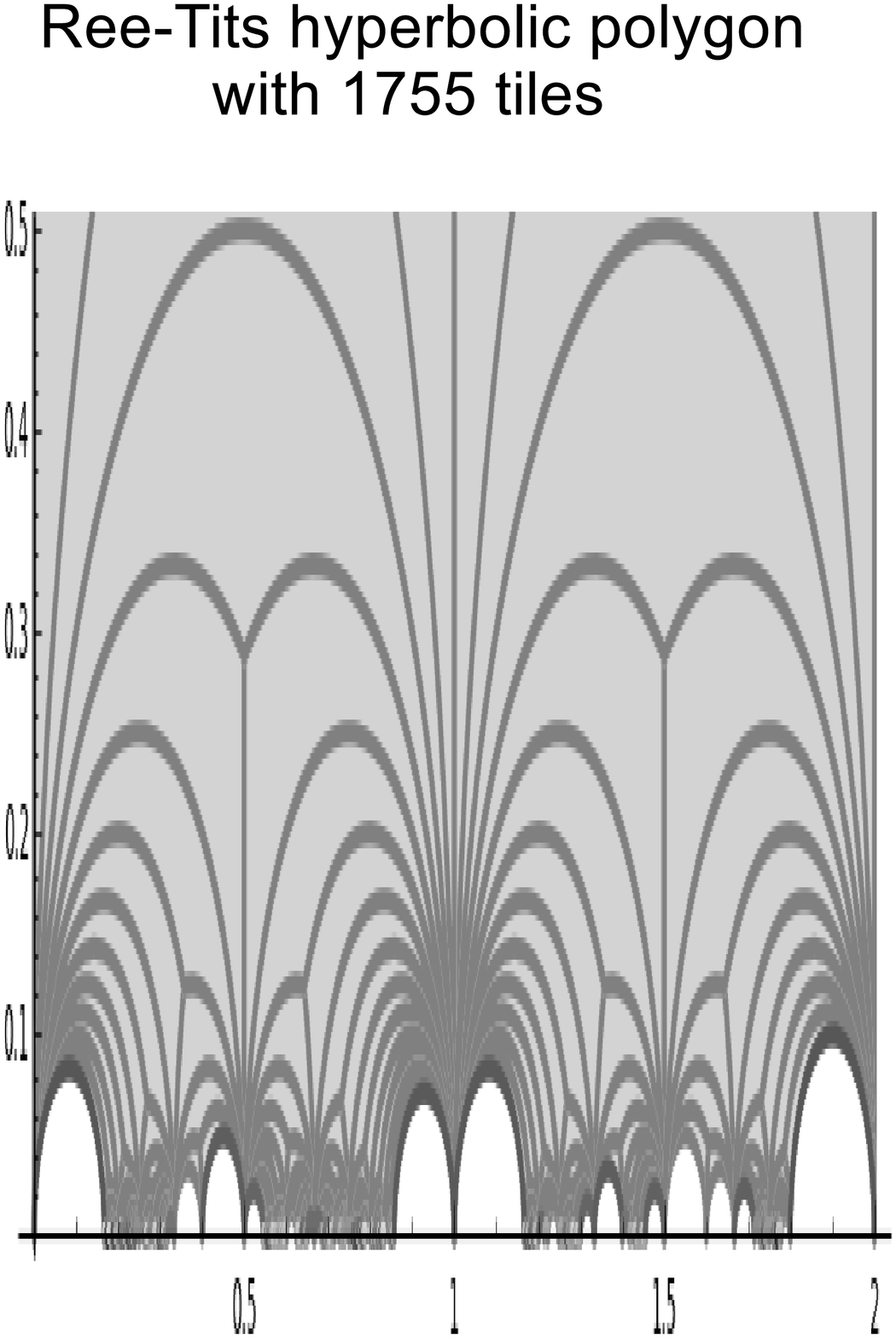}
\includegraphics[width=4cm]{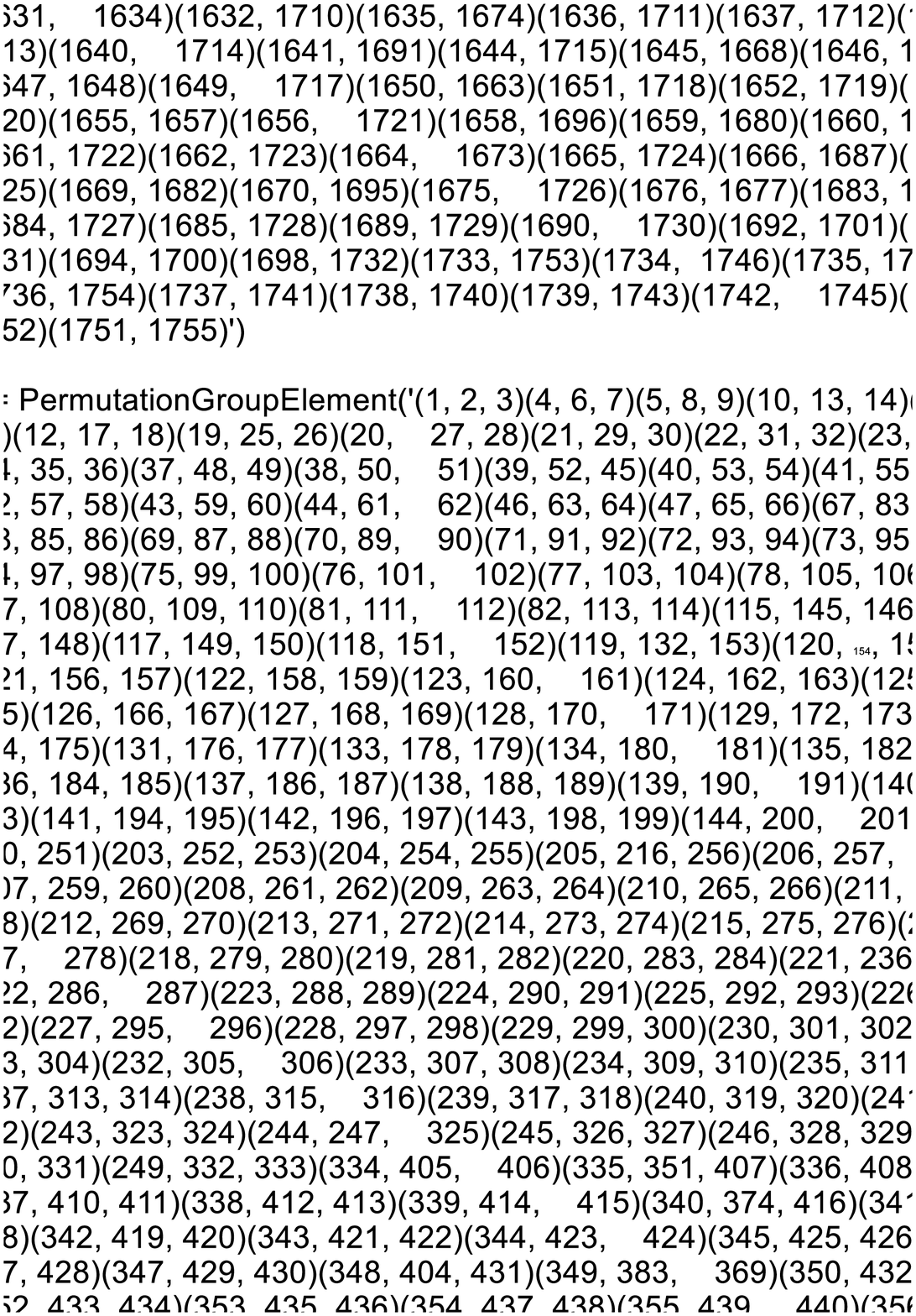}
\caption{The fundamental domain for the index $1755$ permutation group stabilizing the Ree-Tits octagon (details can be found at http://brauer.maths.qmul.ac.uk/Atlas/exc/TF42/mag/TF42G1-p1755B0.M) and an excerpt of the permutation representation.
}
\end{figure}

This is precisely what I did for several \lq sporadic' iguanodons (dessins d'enfants \cite{Groth84},\cite{Dessins2014}). In table 1, I list a few of them (those related to $J_1$, $J_2$, $J_3$, $\mbox{McL}$, $M_{24}$, $T$, $\mbox{Suz}$, $\mbox{Fi}_{23}$, $\mbox{Co}_1$, $\mbox{Fi}'_{24}$)  and give the main characteristics of their fundamental domain $\mathcal{P}$ in the upper-half plane $\mathbb{H}$. All the groups are represented as non-congruence subgroups of the modular group $\Gamma$. The number $n$ of edges of the dessin $\mathbb{D}$ is the index of the (sporadic) group representation.  The genus $g$ of $\mathbb{D}$ equals that of the hyperbolic polygon $\mathcal{P}$, a face of $\mathbb{D}$ corresponds to a cusp of $\mathcal{P}$, the number of black points (resp. of white points) of $\mathbb{D}$ is $B=f+\nu_3-1$ (resp. $W=n+2-2g-B-c$), where $f$ is the number of fractions,  $c$ is the number of cusps, $\nu_2$ and $\nu_3$ are the number of elliptic points of order two and three of $\mathcal{P}$, respectively.

For instance, the smallest permutation representation $P$ of our friend, the Conway group $\mbox{Co}_1$ can be found in the Atlas \cite{Atlasv3}, it may be seen as a dessin $\mathcal{D}$ of index $n=98280$ (half the kissing number of the Leech lattice). An excellent signature consists of the cycle structure \cite{Dessins2014,Lando2004} about the black and white vertices, and the faces of $\mathcal{D}$. For this representation of $\mbox{Co}_1$ one gets the cycles $[3^{32751}1^{27},2^{49140},40^{2400}20^{108}10^{12}]$ \cite{Dessins2014,Lando2004}. From the cycles it is straightforward to recognize that there are $32778=32751+27$ black points (the first of them have valency $3$ and the rest consists of elliptic points of order $3$, i.e. $\nu_3=27$), $49140$ white points (with valency $2$) and $c=2400+108+12=2520$ cusps. The genus $g$ follows from Euler formula given at the beginning of this section. This $\mathcal{D}$ can also be seen as an hyperbolic polygon $\mathcal{P}$ of the upper-half plane (not shown).

Similarly, the dessin $\mathcal{D}$ for the Tits group with permutation representation of index $n=1755$ has the cycle structure $[3^{585}, 2^{8321}1^{91}, 13^{135}]$ and characteristics shown in Table 1. The fundamental domain for this representation of the Tits group $T$ is shown in Fig. 5. The geometry $\mathcal{G}$ stabilized by $\mathcal{D}$  is the Ree-Tits octagon $\mbox{GO}(2,4)$ 
(see https://en.wikipedia.org/wiki/Generalized$\_$polygon for details about the generalized polygons).

\begin{table}[ht]
\begin{center}
\caption{Characteristics of a few small 'sporadic' fundamental polygons.}
% \vspace*{-0.4cm}
\begin{tabular}{||lr|crcl|lr|}
\hline \hline
Graph/geometry& $Group$ & $n$ & $g$  & $\nu_2$ & $\nu_3$ & cusps  & $f$\\
\hline\hline
%\vspace*{-.30cm}
%&&&&&&&&\\
Hall-Janko & $J_2$ &  100 &  0 & 0 &  4 & $1^2 7^{14}$ & 33\\
Livingstone & $J_1$ &  266 &  0 & 10 &  5 & $ 7^{38}$ & 88\\
McLaughin & $McL:2$ &  275&  10 & 11 &  14 & $11^1 22^{12}$ &88\\
$T(24)$ & $M_{24}$ &  276&  10 & 12 &  15 & $23^{12}$ &12\\
Ree-Tits $GO(2,4)$ & T &  1755&  57 & 91 &  0 & $13^{135}$ &586\\
Suzuki & $\mbox{Suz}$ &  1782&  70 & 42 &  0 & $1^1 3^{137}$ &595\\
Janko & $J_3$ &  6156&  321 & 76 &  36 & $19^{324}$ &2041\\
Fischer & $\mbox{Fi}_{23}$ &  31671&  1876 & 695 &  27 & $2^1 4^2 7^9 14^{45} 28^{1106}$&10549\\
Conway & $\mbox{Co}_1$ &  98280 &  6922 & 0 &  27 & $10^{12} 20^{108} 40^{2400}$  & 32752\\
Fischer & $\mbox{Fi}_{24}'$ &  306936&  19409 & 3512 &  0
 & $29^{10584}$&155225\\
\hline
\hline
\end{tabular}
\end{center}
\end{table}

Math: Excellent, the next step would be to compute the modular symbols, the relative homology of the extended upper half plane and the corresponding modular forms of weight two and higher. I am curious to see if the noncongruence cusp forms for these \lq sporadic' polygons have unbounded denominators, as conjectured \cite{LiandLingLong}.

You are close to the \lq philosophy of cusp forms' of Harish-Chandra. He was studying under Dirac himself but turned to mathematics when he learned in  Princeton that {\it not every function is analytic} \cite{Harish2011}. The keyword for the link between moonshine and the Langlands program  is VOA (for vertex operator algebra). The Monster vertex algebra is conjectured (by E. Frenkel and collaborators) to be the unique holomorphic VOA with charge $24$ and partition function $j-744$ (where $j$ is again the modular invariant). 

I quote E. Frenkel \cite{Frenkel}:
{\it Mathematics is not about studying boring and useless equations: It is about accessing a new way of thinking and understanding reality at a deeper level. It endows us with an extra sense and enables humanity to keep pushing the boundaries of the unknown.}  

I suggest that you look at a promising line of moonshine research by I. Frenkel (\lq not E. Frenkel') and J. Dunkan \cite{Dunkan2012} based on the use of a Rademacher sum for the modular invariant $j (\tau)=q^{-1}+744+196884 q +21493760 q^2+\cdots$, with $q=e^{2i\pi \tau}$ as 
$$j (\tau)+12=e^{-2i\pi\tau}+\lim_{K\rightarrow \infty}\sum_{\begin{array}{c}
0<c<K,~ -K^2<d<K^2 \\
(c,d)=1 \end{array}
}e^{-2i\pi\frac{a\tau+b}{c\tau+d}}-e^{-2i\pi\frac{a}{c}},$$
with $a,b$ relative integers and $ad-bc=1$, $\tau \in \mathbb{H}$.

The authors show that the McKay-Thomson series $T_g(\tau)$ of an element $g \in \mathbb{M}$ coincides with a generalized Rademacher sum. They expect to get a version of the simplest chiral $3d$ quantum gravity with $24$ charges, as the original construction of the chiral $2d$ conformal field theory by McKay based on the partition function $j(q)$.

Phys: I am familiar with Rademacher work. Some time ago, I investigated the thermodynamics of the Euler gas whose partition function is that of the number of unrestricted partitions $p(n)$. This was useful to model the low frequency fluctuations (of the $1/f$ type) occurring in a gaz of bosons, like the phonons in a quartz crystal resonator \cite{Planat1fnoise}. The mathematics involves the Hardy-Ramanujan circle method in analytic number theory and this was improved by Rademacher.

Math: Do you think that our mathematics is the real world?
% Or may be we are living in a computer simulation, as proposed by Nick Bostrom. Or our consciousness is a state of matter, as recently discussed by Max Tegmark.

%Then in a recent paper
%{\it In fact the original Rademacher sum does appear as a saddle point approximation to the partition function of the simplest chiral three dimensional quantum gravity}.
Phys: As a provisional response, I offer you a quote of Stephen Hawking from his lecture \lq\lq Godel and the end of the universe" \cite{Hawking}: {\it In the standard positivist approach to the philosophy of science, physical theories live rent free in a Platonic heaven of ideal mathematical models...
% That is, a model can be arbitrarily detailed and can contain an arbitrary amount of information without affecting the universes they describe.
 But we are not angels, who view the universe from the outside. Instead, we and our models are both part of the universe we are describing. Thus a physical theory is self referencing, like in Godel’s theorem. One might therefore expect it to be either inconsistent or incomplete. The theories we have so far are both inconsistent and incomplete.}

Math:
% Yes, self reference, consciousness and contextuality are isomorphic at the bottom and a quote attributed to Darwin is: {\it A mathematician is a blind man in a dark room looking for a black cat that isn't there} \cite{Darwin}.
But you should know Dyson's words in 1981: {\it I have a sneaking hope, a hope unsupported by any facts or any evidence, that sometime in the twenty-first century physicists will stumble upon the Monster group, built in some unsuspected way into the structure of the universe. This is of course only a wild speculation, almost certainly wrong} \cite{Dyson}. 

%\newpage
\tiny

\end{document}